\documentclass[11pt]{article}

\usepackage{bbm}
\usepackage{graphicx}
\usepackage{pdfpages}

\begin{document}

\title{Universal Quantum Algorithm}         
\author{Avatar Tulsi\\
        {\small Department of Physics, IIT Bombay, Mumbai - 400 076, India} \\ {\small tulsi9@gmail.com}}

\maketitle

\begin{abstract}
Quantum amplitude amplification and quantum phase estimation are two fundamental quantum algorithms. All known quantum algorithms are derived from these two algorithms. Even the adiabatic quantum algorithms can also be efficiently simulated using quantum phase estimation. We present a universal quantum algorithm which explains these two algorithms as special cases. An interesting result is that we do not need quantum fourier transform to do quantum phase estimation. 	
\end{abstract}

We present a universal quantum algorithm. We show that quantum phase estimation (see Section $5.2$ of ~\cite{phase}) and quantum amplitude amplification~\cite{grover1,grover2,grover3} are special cases of our universal algorithm. As these two quantum algorithms are known as fundamental quantum algorithms, all other quantum algorithms (including adiabatic quantum algorithms) can be constructed using them. Thus our universal algorithm explains all known quantum algorithms as its special cases. This may help us in finding new quantum algorithms.

We consider an $N$-dimensional quantum system and denote its computational basis states by $|i\rangle$ where $i \in \{0,\ldots,N-1\}$. The initial state $|s\rangle$ can be expanded as $\sum_{i}s_{i}|i\rangle$ where $s_{i}$ are chosen to be real. The universal algorithm is multiple iterations of the operator $U$ on $|s\rangle$ where $U$ is $I_{s}G$. Here $I_{s}$ is the selective phase inversion of $|s\rangle$ given by $\mathbbm{1}- 2|s\rangle \langle s|$ where $\mathbbm{1}$ is the identity operator. The diagonal operator $G$ is defined by its diagonal elements $\langle i|G|i\rangle$ which are $e^{\imath \theta_{i}}$. Up to an ignorable global phase factor, $\theta_{i}$ can be chosen to be zero for a unique basis state which we refer here as the \emph{target} state $|t\rangle$. We desire evolution of quantum system to $|t\rangle$. Thus 
\begin{equation}
|\theta_{i \neq t}| \geq \theta_{\rm min},\ \  \theta_{i = t} = 0. \label{thetas}
\end{equation} 
Here $\theta_{\rm min}$ is the spectral gap of $G$.
  
	The dynamics of universal algorithm can be understood by analyzing the eigenspectrum of $U$. Suppose $U|\lambda\rangle$ is $e^{\imath \lambda}|\lambda\rangle$. Left multiplication by $I_{s}$ implies that $G|\lambda\rangle$ is $e^{\imath \lambda}I_{s}|\lambda\rangle$ as $U$ is $I_{s}G$ and $I_{s}I_{s}$ is $\mathbbm{1}$. Another left multiplication by $\langle i|$ gives us
\begin{equation}
e^{\imath (\theta_{i}-\lambda)}\langle i|\lambda\rangle = \langle i|\lambda\rangle - 2s_{i}\langle s|\lambda\rangle.
\end{equation}
A simple rearrangement of the terms gives us
\begin{equation}
\langle i|\lambda\rangle = \frac{2s_{i}\langle s|\lambda\rangle}{1-e^{\imath (\theta_{i}-\lambda)}} = \imath \frac{s_{i}\langle s|\lambda\rangle e^{-\imath (\theta_{i} - \lambda)/2}} {\sin\frac{\theta_{i}-\lambda}{2}}. \label{equationA}
\end{equation}  
Multiplying by $s_{i}$ and then summing over $i$, we get
\begin{equation}
\sum_{i}s_{i}^{2}\cot\frac{\lambda - \theta_{i}}{2} = 0.
\end{equation} 
which must be satisfied by $\lambda$ for $e^{\imath \lambda}$ to be an eigenvalue of $U$. Generally it is hard to solve but it is simple if we assume $|\lambda| \ll \theta_{\rm min}$ and $s_{t} \ll 1$. Then using the Laurent series expansions for $\cot\frac{\lambda - \theta_{i}}{2}$ and ignoring $O(s_{t}^{2},\lambda^{2})$ terms, above equation becomes 
\begin{equation}
\left(1 + \Lambda_{2}\right)\lambda^{2} + 2\Lambda_{1}\lambda - 4s_{t}^{2} = 0, \label{quadratic}
\end{equation}
where 
\begin{equation}
\Lambda_{p} = \sum_{i \neq t} s_{i}^{2}\cot^{p}\frac{\theta_{i}}{2},\ \ p \in \{1,2\}. \label{momentdefine}
\end{equation}
The product and sum of the two solutions $\lambda_{\pm}$ of Eq. (\ref{quadratic}) are $-4s_{t}^{2}/B^{2}$ and $-2\Lambda_{1}/B^{2}$ respectively. So 
\begin{equation}
\lambda_{\pm} = \pm \frac{2s_{t}}{B}(\tan \eta)^{\pm 1},\ \ \cot 2\eta = \frac{\Lambda_{1}}{2s_{t}B},  \label{eigenvalues}
\end{equation}
where $B$ is $\sqrt{1+\Lambda_{2}}$.

	Now we show that the eigenphases $\lambda_{\pm}$ are the only relevant eigenphases for our purpose as the desired target state $|t\rangle$ is completely spanned by the corresponding eigenstates $|\lambda_{\pm}\rangle$. First, we find $\langle s|\lambda_{\pm}\rangle$ choosen to be real and positive. Putting Eq. (\ref{equationA}) in the normalization condition $\sum_{i}|\langle i|\lambda_{\pm}\rangle|^{2} = 1$ implies that
\begin{equation}
\langle s|\lambda_{\pm}\rangle^{-2} = \sum_{i}s_{i}^{2}\csc^{2}\frac{\lambda_{\pm} - \theta_{i}}{2}.
\end{equation}  
The assumption $|\lambda_{\pm}| \ll \theta_{\rm min} < 1$ implies that in the sum over $i$, the $i = t$ term contributes $4s_{t}^{2}/\lambda_{\pm}^{2}$ which is $B^{2}(\tan \eta)^{\mp 2}$ as per Eq. (\ref{eigenvalues}). Also, using the Laurent series expansions for $\csc\frac{\lambda_{\pm}-\theta_{i}}{2}$, we find that all other $i \neq t$ terms contribute $B^{2}$ if we ignore small $O(\lambda_{\pm}/\theta_{\rm min})$ terms. Thus 
\begin{equation} 
\langle s |\lambda_{\pm}\rangle = B^{-1}f_{\pm}(\eta),\ \ \ f_{\pm}(\eta) = \left(1+\tan^{\mp 2}\eta\right)^{-1/2}. \label{slambdapm} 
\end {equation} 
It is easy to show that $f_{+}(\eta)$ is $\sin \eta$ whereas $f_{-}(\eta)$ is $\cos \eta$. As $\langle s|\lambda_{\pm}\rangle $ are chosen to be real and positive, $\eta $ satisfies $\eta \in [0, \pi/2]$.

	To find $\langle t|\lambda_{\pm}\rangle$, we note that $\theta_{t}$ is zero. Thus, putting Eqs. (\ref{slambdapm}), (\ref{eigenvalues}) and the assumption $|\lambda_{\pm}| \ll \theta_{\rm min}$ in Eq. (\ref{equationA}), we get
\begin{equation}
\langle t|\lambda_{\pm}\rangle = \pm \imath e^{\imath \lambda_{\pm}/2}\frac{f_{\pm}(\eta)}{\tan^{\pm 1} \eta} = \pm \imath e^{\imath \lambda_{\pm}/2} f_{\mp} (\eta). \label{tlambdapm} 
\end{equation}  
As $f_{+}(\eta)^{2}+f_{-}(\eta)^{2}$ is $1$, the desired target state $|t\rangle$ is completely spanned by the eigenstates $|\lambda_{\pm}\rangle$ making them the only relevant eigenstates for our purpose.

	We wish to compute $\alpha(q)$ which is $\langle t|U^{q}|s\rangle$. This can be expanded as $\sum_{j} \langle t|\lambda_{j}\rangle \langle \lambda_{j}|s\rangle e^{\imath q \lambda_{j}}$. As $\langle t|\lambda_{j}\rangle$ is zero for $j \neq \pm$, Eqs. (\ref{slambdapm}) and (\ref{tlambdapm}) alongwith the identity $2f_{\pm} (\eta) f_{\mp} (\eta) = \sin 2\eta$ imply that
\begin{equation}
|\alpha(q)| = \frac{\sin 2\eta}{2B} \left|e^{\imath q'\lambda_{+}}- e^{\imath q'\lambda_{-}}\right| = \frac{\sin 2\eta}{B} \sin \frac{2q's_{t}}{B\sin 2\eta}, \label{alphaq}  
\end{equation}
where $q'$ is $q+\frac{1}{2}$. To get above equation, we have used the fact that $|e^{\imath a} - e^{\imath b}|$ is $2\sin\frac{a-b}{2}$ and $\lambda_{+} - \lambda_{-}$ is $\frac{4s_{t}}{B\sin 2\eta}$ as per Eq. (\ref{eigenvalues}). 

	We choose $q$ to be 
\begin{equation}
q = Q = \frac{\pi B}{4s_{t}} - \frac{1}{2}. \label{query}
\end{equation}
Then Eq. (\ref{alphaq}) implies that the success probability of getting the desired target state $|t\rangle$ is
\begin{equation}
P = |\alpha(Q)|^{2} = \frac{\sin^{2}2\eta}{B^{2}}\sin^{2}\left(\frac{\pi}{2\sin 2\eta}\right). \label{success}
\end{equation}
As $B \geq 1$, $P$ is very small if $\sin 2\eta \ll 1$ which happens when $\cot 2\eta \gg 1$. To keep $P$ large enough, we assume that $\cot 2\eta$ is small. When $\cot 2\eta$ is zero then $\eta$ is $\frac{\pi}{4}$. So, for small values of $\cot 2\eta$, $\eta$ is close to $\frac{\pi}{4}$ or $\Delta$ is small where $\Delta$ is defined as $\eta = \frac{\pi}{4}  + \Delta$. Thus $\cot 2\eta$ is $-\tan 2\Delta \approx -2\Delta$ for small $\Delta$. Eq. (\ref{eigenvalues}) is rewritten as
\begin{equation}
\Delta = -\Lambda_{1}/(4s_{t}B) \ll 1.  \label{Deltavalue}
\end{equation}  
Also, $\sin^{2}2\eta$ is $1-4\Delta^{2}$ and $\sin^{2}\left(\frac{\pi}{2\sin 2\eta}\right)$ is $1-\pi^{2}\Delta^{4}$ so that $P$ is $\Theta(1/B)$. Thus $\Theta(B^{2})$ times preparations of $|\alpha(Q)\rangle$ state are needed to get the target state $|t\rangle$. Each preparation of $|\alpha(Q)\rangle$ needs $Q = \Theta(B/s_{t})$ applications of the operator $U$ so a total of 
\begin{equation}
\Theta(B^{3}/s_{t}) \label{performance}
\end{equation}
applications of $U$ are needed for the success of our algorithm. Also, for small $\Delta$, $\tan \eta$ is $\tan \left(\frac{\pi}{4} + \Delta\right)$ or $1+2\Delta$ so Eq. (\ref{eigenvalues}) implies that the assumption $|\lambda_{\pm}| \ll \theta_{\rm min}$ for the validity of our analysis is equivalent to 
\begin{equation}
\left(2s_{t}/B\right)(1 \pm \Delta) \ll \theta_{\rm min}. \label{validity}
\end{equation}
This completes the analysis of the universal algorithm.

	The eigenvalues of $U$ are evaluated with an error of $O(s_{t}^{2},\lambda_{\pm}^{2})$. So the analysis holds for $q_{\rm max}$ iterations of $U$ with an error of $q_{\rm max}O(s_{t}^{2},\lambda_{\pm}^{2})$. Typically $\lambda_{\pm}$ is $\Theta(s_{t})$ so that $q_{\rm max}$ is $\Theta (s_{t}^{-1})$. Thus the total error is $\Theta(s_{t})$ which is very small. The eigenstates of $U$ are evaluated with an error of $O(\lambda_{\pm}/\theta_{\rm min})$ which is small due to the assumption $|\lambda_{\pm}| \ll \theta_{\rm min}$.

	\emph{Quantum Amplitude Amplification}: Consider the special case when $G$ is $-I_{t}$ where $I_{t}$ is the selective phase inversion of the target state $|t\rangle$. Thus $G$ is $2|t\rangle\langle t| - \mathbbm{1}$ and $\theta_{i \neq t}$ is $\pi$ for all $i \neq t$. The moments $\Lambda_{1}$ and $\Lambda_{2}$ are zero so that $\eta$ is $\frac{\pi}{4}$ and $B$ is $1$. Thus $P_{\rm max}$ is $1$ and $q_{\rm max}$ is $\pi/4s_{t}$. This performance of the quantum amplitude amplification is also known to be the best possible performance. In Grover's search algorithm $|s\rangle$ is chosen to be a uniform superposition of all $N$ basis states. Thus $|s\rangle$ is $\sum_{i}|i\rangle /\sqrt{N}$ and $s_{t}$ is $1/\sqrt{N}$. So the Grover's algorithm finds the target state in $O(\sqrt{N})$ time steps which is quadratically faster than classical algorithms which take $O(N)$ time steps.

	\emph{Quantum Phase estimation}: Suppose we are given a copy of an eigenstate $|\phi\rangle$ corresponding to an eigenvalue $e^{\imath 2\pi \phi}$ of a unitary operator $V$. In the phase estimation, the goal is to estimate $\phi$. Let $M$ be $2^{m}$ for an integer $m$. We estimate $\phi$ by finding an integer $b \in \{0,M-1\}$ such that $\frac{b}{M} = 0.b_{1}\ldots b_{m}$ is the nearest $m$-bit approximation to $\phi$. Here $b_{k}$ ($k \in \{1,\ldots,m\}$) denote the bit values of the binary expression of $b$. So $b$ is $\sum_{k = 1}^{m}2^{m-k}b_{k}$. We define $\delta$ as
\begin{equation}
\delta = \phi - (b/M),\ \ |\delta| \leq (1/2M). \label{deltadefine}
\end{equation}
The inequality follows from the definition of $b$.

	Consider a $2M$-dimensional quantum system of $1+m$ qubits. The first qubit controls the crucial operators and we call this the \emph{control}-qubit whose basis states are denoted by $\{|0'\rangle, |1'\rangle\}$. The remaining $m$ qubits provide estimate of $\phi$ and we call them the \emph{estimate}-qubits labelled by an index $k \in \{1,\ldots,m\}$. The joint Hilbert space of all $m$ estimate-qubits has $M = 2^{m}$ basis states. Each basis state $|\ell\rangle$ encodes an integer $\ell$ which is $\sum_{k=1}^{m}2^{m-k}\ell_{k}$ where $\ell_{k} \in \{0,1\}$ denote the bit values of the binary expression of $\ell$. Thus $|\ell\rangle$ is $\prod_{k=1}^{m}|\ell_{k}\rangle$ where $|\ell_{k}\rangle$ denote the basis state of $k^{\rm th}$ estimate-qubit which is $|0\rangle$ or $|1\rangle$ depending upon the value of $\ell_{k}$.  

	The initial state $|s\rangle$ is chosen to be $|\sigma\rangle|+\rangle^{\otimes m}$ or $(1/\sqrt{M})|\sigma\rangle\sum_{\ell}|\ell\rangle$. Thus the control-qubit state is $|\sigma\rangle$ where $\langle \sigma | 0'\rangle$ is $1/(2\sqrt{M})$ and all $m$ estimate-qubits are in $|+\rangle$ states where $|+\rangle$ is $(|0\rangle + |1\rangle)/\sqrt{2}$. The operator $G$ is chosen to be $Z'(c_{1}V)(c_{0}R)$. Here $Z'$ acts on the control-qubit to multiply the $|1'\rangle$ state by a phase of $e^{\imath \pi}$. The controlled operator $c_{1}V$ applies $V$ on its eigenstate $|\phi\rangle$ if and only if the control qubit is in $|1'\rangle$ state. This multiplies the $|1'\rangle$ state by a phase of $e^{\imath 2\pi \phi}$. Another controlled operation $(c_{0}R)$ applies the operator $R$ on $m$ estimate-qubits if and only if the control qubit is in $|0'\rangle$ state. The operator $R$ is such that $R|\ell\rangle$ is $e^{\imath 2\pi \ell/M}|\ell\rangle$ or $\prod_{k=1}^{m} e^{\imath 2\pi 2^{-k}\ell_{k}}|\ell_{k}\rangle$. Thus $R$ is a product of $m$ single qubit gates $R_{k}$ where each $R_{k}$ acts only on $k^{\rm th}$ estimate-qubit to multiply its $|1\rangle$ state by a phase of $e^{\imath 2\pi/2^{k}}$. Hence $c_{0}R$ is a product of $m$ two-qubit gates.

	Multiplying $G$ by an ignorable global phase factor of $e^{-\imath 2\pi b/M}$, we can write it as 
\begin{equation}
G|0\ell\rangle = e^{\imath 2\pi (\ell-b)/M}|0\ell\rangle,\ \ G|1\ell\rangle  = -e^{\imath 2\pi (\phi-(b/M))}|1\ell\rangle = e^{\imath \pi(1+2\delta)}|1\ell \rangle. \label{Gvalue}
\end{equation}
Here $|0\ell\rangle$ and $|1\ell\rangle$ denote the $|0'\rangle|\ell\rangle$ and $|1'\rangle|\ell\rangle$ states respectively. As $G|i\rangle$ is $e^{\imath \theta_{i}}|i\rangle$, we get
\begin{equation}
\theta_{0\ell} = 2\pi(\ell - b)/M,\ \ \theta_{1\ell} = \pi(1+2\delta),\ \ \theta_{\rm min} = 2\pi/M. \label{thetavalues}
\end{equation}
The eigenphase $\theta_{i}$ is zero if and only if $|i\rangle$ is $|0b\rangle$. Thus $|0b\rangle$ is the target state $|t\rangle$ which provides an estimate of $\phi$ in terms of $b$. 

	As $|s\rangle$ is $(1/\sqrt{M})|\sigma\rangle\sum_{\ell}|\ell\rangle$ and $\langle \sigma|0'\rangle$ is $1/2\sqrt{M}$, we find that the coefficients $s_{0\ell}$ are $1/2M$ whereas $s_{1\ell}$ are $(1/\sqrt{M})(1-\Theta(1/M))$. Thus the moments $\Lambda_{p}$ as per Eq. (\ref{momentdefine}) are found to be, ignoring $O(M^{-3/2})$ terms,
\begin{equation}
\Lambda_{p} = (-\pi \delta)^{p} + (4M^{2})^{-1}\sum'_{\ell}h_{\ell}^{p},\ \ h_{\ell} = \cot\frac{\pi}{M}(\ell - b). \label{momentvalue}
\end{equation}
The prime notation indicates that the sum excludes $\ell = b$ term. To evaluate this sum, we note that $h_{\ell}$ is zero when $\ell$ is $b+\frac{M}{2}$. Other integers $\ell \neq \{b,b+(M/2)\}$ come in pairs $\{\ell,\ell'\}$ where $\ell'$ is $(2b-\ell)\rm mod M$ so that $h_{\ell}$ is $-h_{\ell'}$. Thus $\sum'_{\ell}h_{ell}$ is zero and $\sum'_{\ell}h_{ell}^{2}$ is $2\sum''_{\ell} h_{\ell}^{2}$ where the double prime indicates that the sum is over all $\ell$ from $(b+1) \rm mod M$ to $(b+\frac{M}{2}) \rm mod M$. The significant contributions to this sum is due to the terms for which $\pi(\ell - b)/M$ is small and $h_{\ell}$ is $\frac{M}{\pi(\ell - b)}$. Thus this sum is approximately $2\sum_{r = 1}^{M/2}\frac{1}{r^{2}}$ which converges to $\frac{\pi^{2}}{3}$ for large $M$. Thus $\Lambda_{2}$ is $\frac{1}{12}$ and $B = \sqrt{1+\Lambda_{2}}$ is $ 1.04$ or $\Theta(1)$. 

	As $|t\rangle$ is $|0b\rangle$, $s_{t}$ is $1/2M$ and Eq. (\ref{performance}) implies that $b$ can be found using $\Theta(M)$ applications of $U$ which is $I_{s}G$ provided the assumption of small $\Delta$ is correct. As $|\delta| \leq 1/2M$, $\delta$ lies in the interval $\{\frac{-1}{2M},\frac{1}{2M}\}$ or $\{\frac{-4}{8M},\frac{4}{8M}\}$. We can always write $\delta$ as $\frac{g}{8M}+\delta'$ where $|\delta'| \leq \frac{1}{8M}$ and $g \in \{\pm 1,\pm 3\}$. We run our universal algorithm $4$ times where each running correspond to a unique value of $g$. In each running, we use the operator $V' = e^{-\imath 2\pi g/8M}V$ in place of $V$ to construct the operator $G$. The $|\psi\rangle$ is an eigenstate of $V'$ with the eigenvalue $e^{\imath 2\pi \phi'}$ where $\phi'$ is $\phi - \frac{g}{8M}$ or $\frac{b}{M}-\frac{g}{8M}+\delta$ or $\frac{b}{M} + \delta'$. Thus our analysis can be used if we put $\delta'$ in place of $\delta$. Then Eq. (\ref{momentvalue}) implies that $\Lambda_{1}$ is $-\pi \delta'$. Putting this in Eq. (\ref{Deltavalue}), we find that $\Delta$ is $1.51 \delta' M$ as $s_{t}$ is $1/2M$ and $B$ is $1.04$. Then $|\delta'| \leq \frac{1}{8M}$ implies that $|\Delta| \leq 0.19$ and thus $\Delta$ is small enough for our analysis to hold. Note that $4$ runnings of the algorithm also improve the accuracy of estimation of $\phi$ by a factor of $4$ from $\pm \frac{1}{2M}$ to $\pm \frac{1}{8M}$. 

	It is easy to check that $\Delta$ is small only if we choose the \emph{ignored} global phase factor in Eq. (\ref{Gvalue}) to be $e^{-\imath 2\pi b/M}$. Suppose we choose this phase to be $e^{-\imath 2\pi b'/M}$ where $b'$ is an integer far from $b$. Then $\Lambda_{1}$ is such that $|\cot 2\eta| = \Lambda_{1}/2s_{t}B$ is too large making $\sin 2\eta$ too small so that the success probability also becomes very small. The condition $\Lambda_{1} \ll 2s_{t}B$ for the success of our algorithm is quite similar to the phase-matching condition of Grover's algorithm and it is due to resonance phenomena. This plays a crucial part in estimating $\phi$.


\begin{thebibliography}{99}
\bibitem{phase} M. Nielsen and I. Chuang (2010), Quantum Computation and Quantum Information, Cambridge University Press.
\bibitem{grover1} L.K. Grover, Phys. Rev. Lett. \textbf{79}, 325 (1997).
\bibitem{grover2} L.K. Grover, Phys. Rev. Lett. \textbf{80}, 4329 (1998).
\bibitem{grover3}C. Bennett, E. Bernstein, G. Brassard, and U. Vazirani, SIAM J. Computing {\bf 26}, 1510 (1997) [arXiv.org:quant-ph/9701001].
\end{thebibliography}
\end{document}